\documentclass[8pt]{article}
\usepackage{times}

\usepackage{amssymb}

\usepackage[figuresright]{rotating}

\begin{document}






\begin{center}
{\Large \bf Pinning down QCD-matter shear viscosity in ultrarelativistic heavy-ion collisions via EbyE fluctuations using pQCD + saturation + hydrodynamics}
\vspace{0.5cm}



{\large K. J. Eskola$^{a}\footnote{Speaker at the \textit{Hard Probes 2015}, Montreal, Canada, June 2015.}$, H. Niemi$^b$ and R. Paatelainen$^c$}\\
\vspace{0.5cm}


{\small 
$^a$University of Jyvaskyla, Department of Physics, P.O. Box 35, FI-40014 University of Jyvaskyla, Finland\\
$^b$Institut f\"ur Theoretische Physik, J.~W.~Goethe-Universit\"at,
Max-von-Laue-Str. 1, D-60438 Frankfurt am Main, Germany\\
$^c$Departamento de Fisica de Particulas, Universidade de Santiago de Compostela,
E-15782 Santiago de Compostela, Galicia, Spain\\
}
\vspace{0.5cm}

\end{center}


\begin{abstract}
We introduce an event-by-event pQCD + saturation + hydro ("EKRT") framework for high-energy heavy-ion collisions, where we compute the produced fluctuating QCD-matter energy densities from next-to-leading order (NLO) perturbative QCD (pQCD) using saturation to control soft particle production, and describe the space-time evolution of the QCD matter with viscous hydrodynamics, event by event (EbyE). We compare the computed centrality dependence of hadronic multiplicities, $p_T$ spectra and flow coefficients $v_n$ against LHC and RHIC data. We compare also the computed EbyE probability distributions of relative fluctuations of $v_n$, as well as correlations of 2 and 3 event-plane angles, with LHC data. Our systematic multi-energy and -observable analysis not only tests the initial state calculation and applicability of hydrodynamics, but also makes it possible to constrain the temperature dependence of the shear viscosity-to-entropy ratio, $\eta/s(T)$, of QCD matter in its different phases. Remarkably, we can describe all these different flow observables and correlations consistently with $\eta/s(T)$ that is independent of the collision energy.
\end{abstract}





\section{Introduction}
\label{intro}

The basic idea in the new NLO-improved perturbative QCD + saturation + viscous hydro event-by-event EKRT framework \cite{Niemi:2015qia} presented in this talk, is as follows: Using collinear factorization and NLO pQCD, we first compute the production of transverse energy ($E_T$) carried by partons of $p_T\sim$ few GeV (minijets) into the central region in $A$+$A$ collisions. The conjecture of gluon saturation, implemented as in the NLO-extension \cite{Paatelainen:2012at} of the original EKRT paper \cite{Eskola:1999fc}, then allows us to compute QCD-matter initial conditions, EbyE, for viscous hydrodynamics with which we describe the space-time evolution of the produced QCD matter. Comparing with a multitude of experimental RHIC and LHC data for low-$p_T$ observables, we aim at \textit{(i)} pinning down the $\eta/s(T)$ ratio,
\textit{(ii)} testing our initial state calculation, and \textit{(iii)} studying the applicability region of viscous hydrodynamics.

\section{Minijet $E_T$ production from NLO pQCD}
\label{minijets}

Using collinear factorization, NLO minijet $E_T$ production per transverse area into a rapidity interval $\Delta y$ in high-energy $A$+$A$ collisions of impact parameter 
$\mathbf{b}$  can be computed as  
\begin{equation}
\frac{\mathrm{d}E_T}{\mathrm{d}^2\mathbf{r}} = T_A(\mathbf{r}+ \mathbf{b}/2)T_A(\mathbf{r}- \mathbf{b}/2) \sigma\langle E_T\rangle, 
\label{eq:ET}
\end{equation}
where the collision geometry is given by the the standard nuclear thickness functions $T_A$. The minijet part \cite{Eskola:1988yh}, which in NLO \cite{Eskola:2000my,Paatelainen:2012at,Niemi:2015qia} can be written as
\begin{equation}
\label{eq:sigmaET2}
\sigma\langle E_T\rangle =  \sum_{n=2}^{3}\frac{1}{n!}\int \mathrm{d[PS]}_n\frac{\mathrm{d}\sigma^{2\rightarrow n}}{\mathrm{d[PS]}_n}\tilde{\mathcal{S}}_n,
\end{equation}
contains phase-space integrated $2\rightarrow2$ and $2\rightarrow3$ differential ${\cal O}(\alpha_s^3)$ partonic cross sections $\mathrm{d}\sigma^{2\rightarrow n}/\mathrm{d[PS]}_{n}$, obtained in terms of UV-renormalized squared matrix elements \cite{Ellis:1985er,Paatelainen:2014fsa}, and 
CTEQ6M \cite{Pumplin:2002vw} parton distribution functions with EPS09s spatially dependent nuclear effects \cite{Helenius:2012wd}. The measurement functions,
\begin{eqnarray}
\label{eq:mfunctions}
\tilde{\mathcal{S}}_n &=& 
\left(\sum_{i=1}^{n} \theta(y_i \in \Delta y)p_{Ti}\right) 
\times 
\theta \left (\sum_{i=1}^{n}p_{Ti}  \geq 2p_0 \right ) 
\times 
\theta\left(\sum_{i=1}^{n} \theta(y_i \in \Delta y)p_{Ti} \geq \beta p_0\right), 
\end{eqnarray}
contain step functions $\theta$ and are analogous to those in jet production \cite{Kunszt:1992tn}. They define the
1) minijet $E_T$ as the sum of the minijet $p_T$'s in $\Delta y$; 
2) $p_T$ cut-off scale $p_0\gg\Lambda_{\rm QCD}$ above which we do the minijet calculation, and 
3) minimum $E_T$ that we might require in the interval $\Delta y$. Such $\tilde{\mathcal{S}}_n $ ensure a well-defined NLO calculation safe from collinear and infra-red singularities. With these ingredients, the minijet $E_T$ calculation is field-theoretically rigorous for fixed $p_0$ and $\beta\in[0,1]$. As the $E_T$ here is not a direct observable, we must leave the parameter $\beta$ to be fixed from experimental data.

\section{Saturation in NLO minijet production}
\label{saturation}

The parameter $p_0$  controlling minijet production is here fixed based on a gluon saturation conjecture. As argued in  \cite{Paatelainen:2012at}, the $E_T$ production is expected to cease (saturate) when $3\rightarrow2$ and higher order processes  become of the same magnitude as the basic $2\rightarrow2$ ones, $\label{eq:ETsat}
\frac{\mathrm{d}E_T}{\mathrm{d}^2\mathbf{r}\mathrm{d}y}(2\rightarrow 2)  \sim
\frac{\mathrm{d}E_T}{\mathrm{d}^2\mathbf{r}\mathrm{d}y}(3\rightarrow 2)$.
This leads to a saturation criterion analogous to that in the original EKRT-model \cite{Eskola:1999fc} but now for an IR/CL-safe $E_T$,
\begin{equation}
\label{eq:satcri}
\frac{\mathrm{d}E_T}{\mathrm{d}^2\mathbf{r}}(p_0,\sqrt{s_{NN}},A,\Delta y,\mathbf{r},\mathbf{b};\beta) = \frac{K_{\rm sat}}{\pi}p_0^3\Delta y,
\end{equation}
from which $p_0 = p_{{\rm sat}}(\sqrt{s_{NN}},A,\Delta y,\mathbf{r},\mathbf{b};\beta,K_{\rm sat})$ is solved locally in {\bf r} and for fixed $\beta,K_{\rm sat}$. The "packing factor", proportionality constant $K_{\rm sat}$, is to be fixed from the data. The key observation \cite{Paatelainen:2013eea,Eskola:2001rx} enabling the EbyE framework here, is that $p_{\rm{sat}}$ scales essentially as $[T_AT_A]^n$, 
so that we can parametrize  $p_{\rm sat }(T_AT_A)$ vs. $\beta,K_{\rm sat}$. The parametrization, intended also for public use, can be found in \cite{Niemi:2015qia}. The minimum $p_{\rm sat}$ we allow here is 1 GeV.

\section{NLO EKRT EbyE framework}
\label{EbyE_EKRT}
For the EbyE set-up, we first sample the nucleon positions in the colliding nuclei from the standard two-parameter Woods-Saxon density. Around each nucleon, we then set a gluon transverse density, 
$T_{n}(r)=(2\pi\sigma^{2})^{-1}\exp(-r^2/(2\sigma^2))$,
whose width parameter, $\sigma= 0.43$~fm is obtained from the exclusive electroproduction data of $J/\psi$
at HERA/ZEUS \cite{Chekanov:2004mw}.
After this the thickness functions $T_A(x,y)$ of the nuclei can be computed, and $p_{\rm sat}(T_AT_A)$ obtained for fixed $\beta,K_{\rm sat}$ on the basis of the NLO pQCD+saturation calculation described above. From $p_{\rm sat}$, we then extract the local formation time, $\tau_s(\mathbf{r}) = 1/p_{\rm sat}(\mathbf{r})$, and energy density
\begin{equation}
e(\mathbf{r},\tau_{\mathrm{s}}(\mathbf{r})) = \frac{\mathrm{d}E_T}{\mathrm{d}^2\mathbf{r}}\frac{1}{\tau_{\mathrm{s}} (\mathbf{r}) \Delta y } = \frac{K_{\rm sat}}{\pi}[p_{\rm sat}(\mathbf{r})]^4
\end{equation}
of the minijet plasma.

"Pre-thermal" evolution from $\tau_s(\mathbf{r})$ to $\tau_0 = 1/p_{\rm sat}^{\rm min}=0.2$~fm is here done simply with 1 D Bjorken hydro (also free streaming would work \cite{Paatelainen:2013eea}). At the edges of the collision system, below $p_{\rm sat}^{\rm min}=1$ GeV, we connect the $e$-profile smoothly to a binary profile (see \cite{Niemi:2015qia}).

\begin{figure}
\begin{center}
\includegraphics[width=7.3cm]{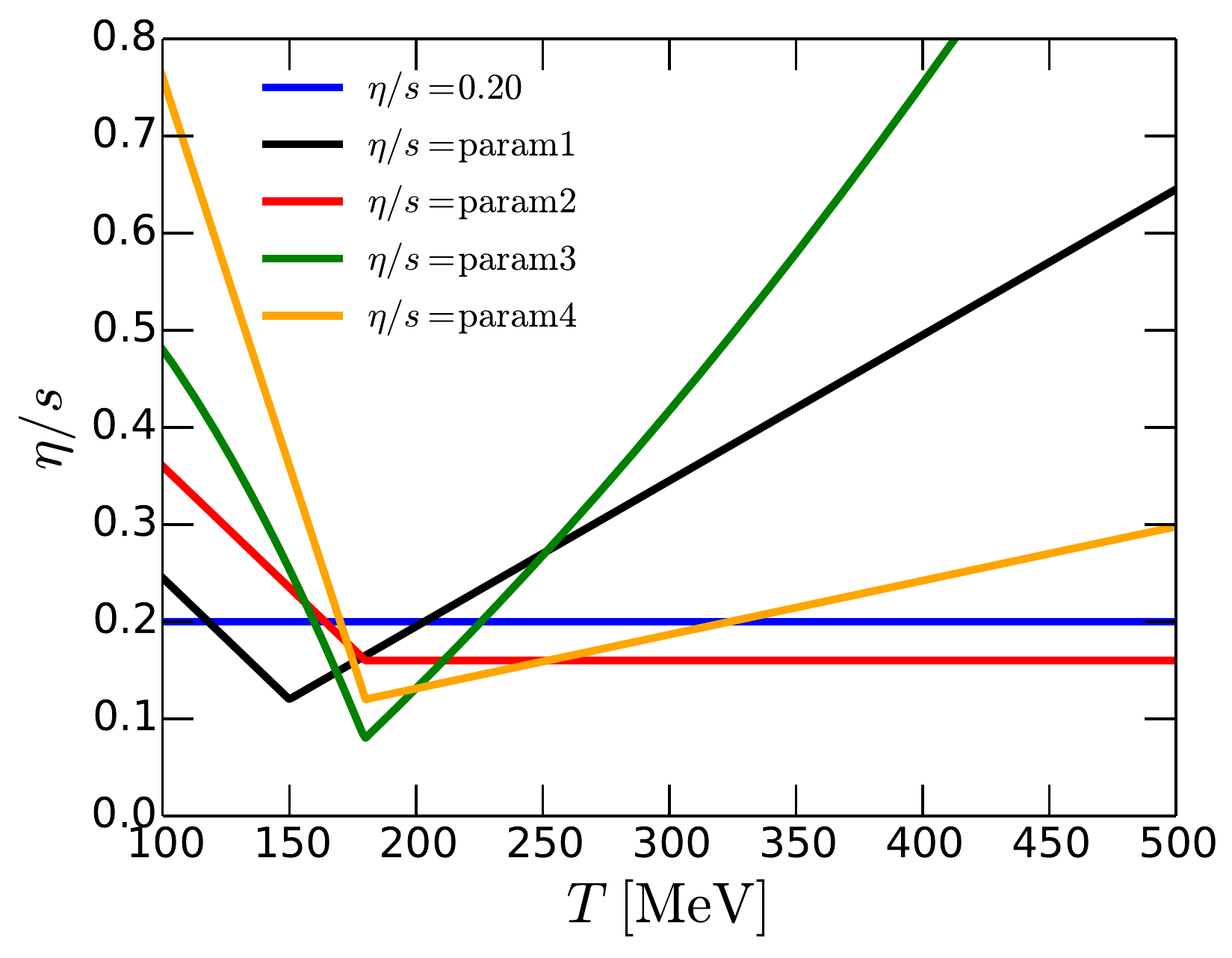}
\end{center}
\vspace{-0.5cm}
\caption{\small The tested temperature dependences of $\eta/s$. From \cite{Niemi:2015qia}.}
\label{fig:etapers}
\end{figure}

With such EbyE initial conditions, we then run 2+1 D viscous hydrodynamics, EbyE. Our 
hydro setup is the one used in \cite{Niemi:2011ix, Niemi:2012ry, Niemi:2012aj, Paatelainen:2013eea}, i.e. 2nd-order dissipative relativistic hydro with transient fluid-dynamics equation of motion for the shear-stress tensor $\pi^{\mu\nu}$ \cite{Denicol:2010xn, Denicol:2012cn, Molnar:2013lta} (see again \cite{Niemi:2015qia}). We neglect heat conductivity and bulk viscosity and study the effects of shear viscosity only. We explore the possible temperature dependence of $\eta/s(T)$ with the parametrizations shown in Fig.~\ref{fig:etapers}.

As the QCD-matter equation of state, we employ the $s95p$-PCE-v1 parametrization \cite{Huovinen:2009yb} with chemical decoupling at a rather high temperature of 175 MeV, and kinetic freeze-out at 100 MeV. Resonance decays after the freeze-out are included. The initial $\pi^{\mu\nu}(\tau_0)$ and transverse flow $v_T(\tau_0)$ are set to zero. In the non-equilibrium particle distributions on the freeze-out surface, we assume, as usual, that the relative deviations from the equilibrium distributions are proportional to $p_\mu p_\nu\pi^{\mu\nu}$ for each particle species.

\section{Comparison with LHC and RHIC data}
\label{data}
\begin{figure*}
\vspace{-0.5cm}
\begin{center}
\includegraphics[width=6.5cm]{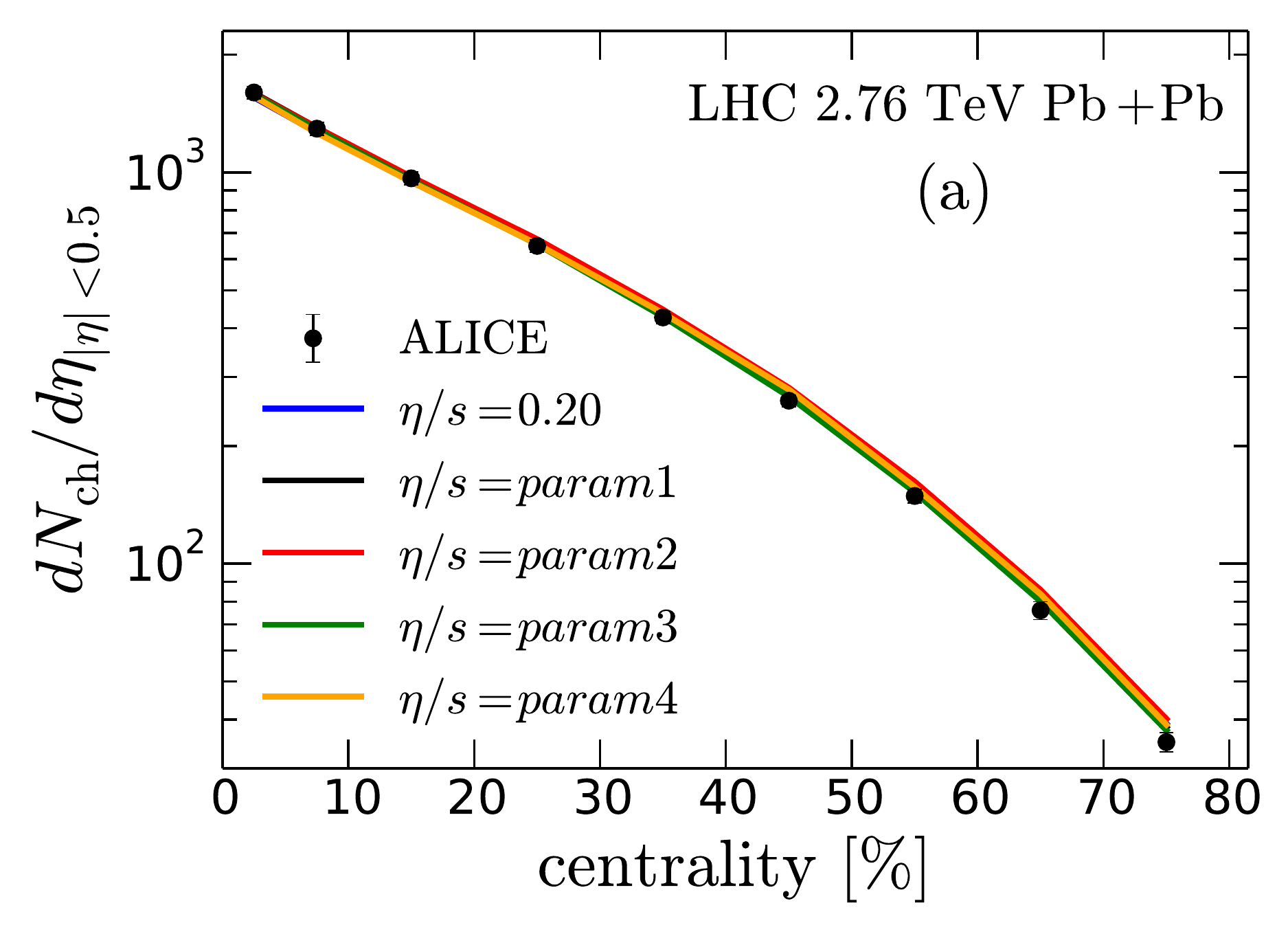}
\includegraphics[width=6.5cm]{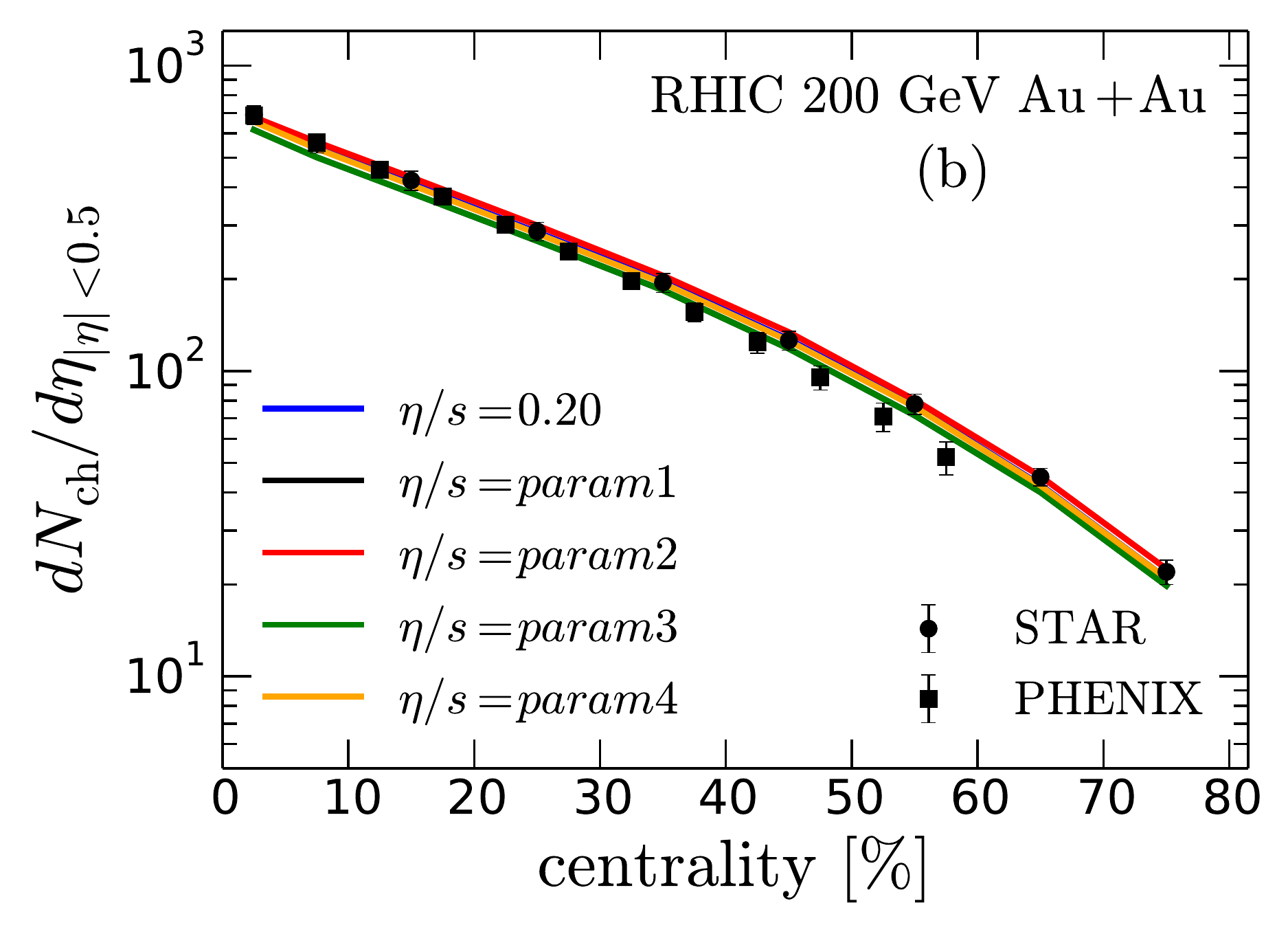} \\
\vspace{-0.2cm}
\includegraphics[width=6.5cm]{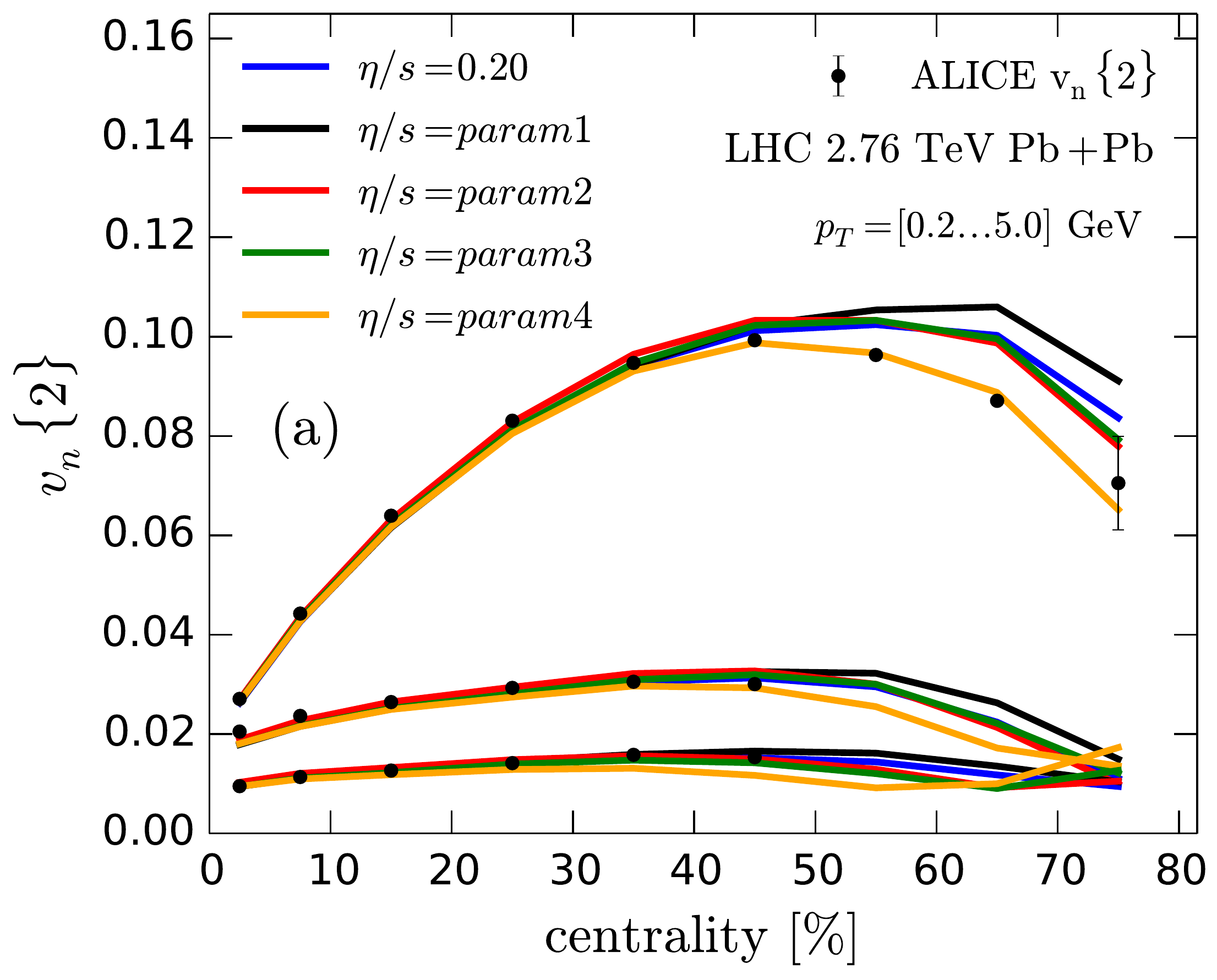} 
\includegraphics[width=6.6cm]{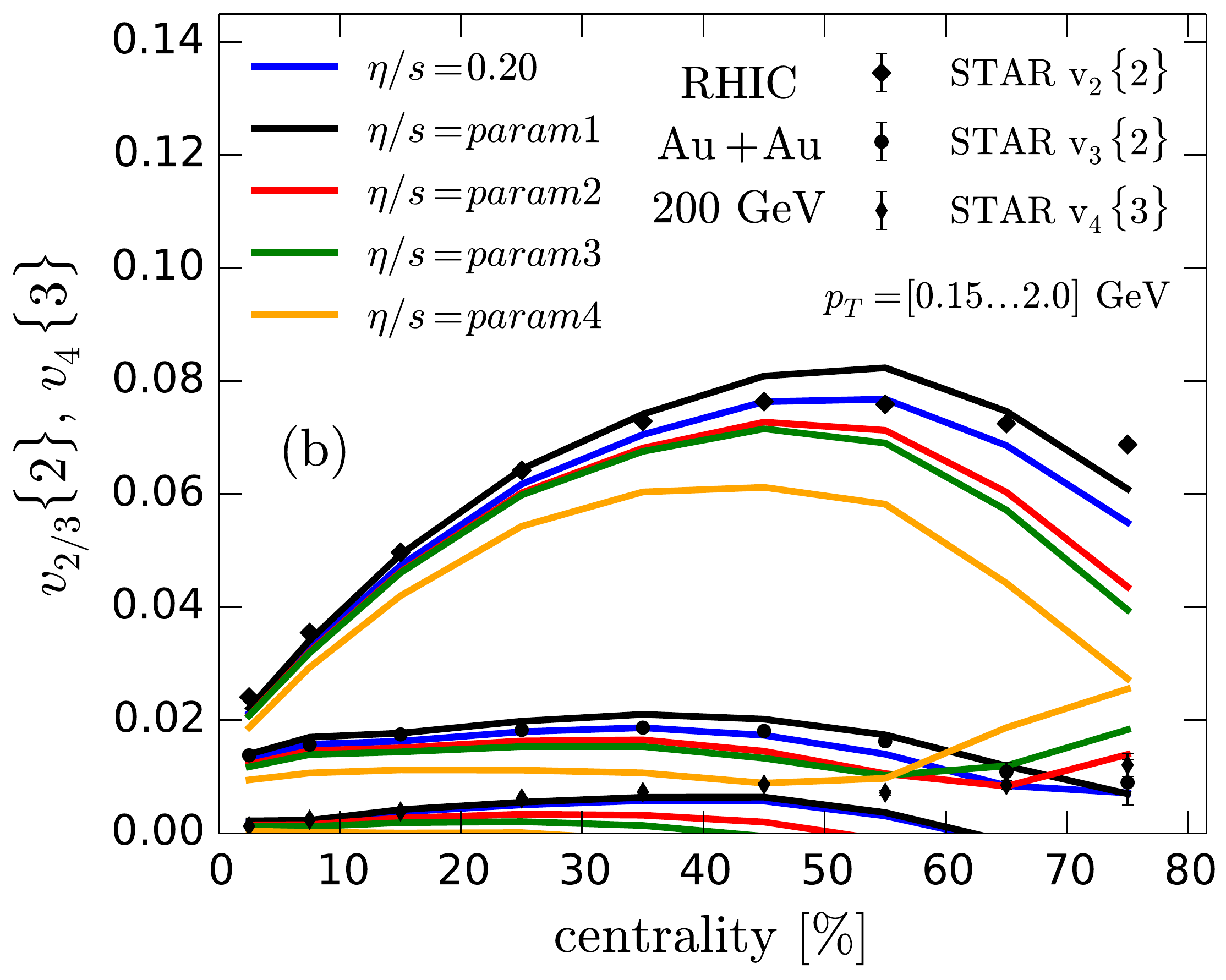}
\end{center}
\vspace{-0.5cm}
\caption{\small \textit{Upper panels:} Centrality dependence of charged hadron multiplicities at the LHC (a) and RHIC (b). \textit{Lower panels:} Centrality dependence of the 2-hadron cumulant flow coefficients $v_n\{2\}$ at the LHC (a), and $v_2\{2\}$, $v_3\{2\}$, $v_4\{3\}$ from the charged hadron 2- and 3-particle cumulants at RHIC (b). 
Experimental multiplicity data are from \cite{Aamodt:2010cz} (a) and \cite{Abelev:2008ab,Adler:2004zn} (b), and $v_n$ data from \cite{ALICE:2011ab} (a) and \cite{Adams:2004bi, Adamczyk:2013waa, Adams:2003zg} (b). From \cite{Niemi:2015qia}. 
}
\label{fig:LHC_RHIC_data}
\vspace{-0.0cm}
\end{figure*}

Let us then consider selected results from the extensive multi-energy and multi-observable analysis of Ref.~\cite{Niemi:2015qia}. Figure \ref{fig:LHC_RHIC_data} 
shows the centrality dependence of the computed charged hadron multiplicity $dN_{ch}/d\eta$ and their flow coefficients $v_n\{2\}$ from the 2-particle cumulants at the LHC and $v_{2,3}\{2\}, v_{4}\{3\}$ at RHIC, and comparison with the data. For details, definitions of these quantities, and $p_T$ spectra, see \cite{Niemi:2015qia}. For each $\eta/s(T)$ parametrization, and setting $\beta=0.8$ \cite{Paatelainen:2013eea}, we exploit only the centralmost LHC multiplicity datapoint to fix the normalization with $K_{\rm sat}$. The good agreement with the data indicates that our initial states are on the average working well. All the $\eta/s(T)$ parametrizations of Fig.~\ref{fig:etapers} reproduce the measured $v_n\{2\}$'s at the LHC while the RHIC data favors a small $\eta/s(T)$ in the hadron gas phase. 

\begin{figure*}
\vspace{-0.0cm}
\includegraphics[width=5.5cm]{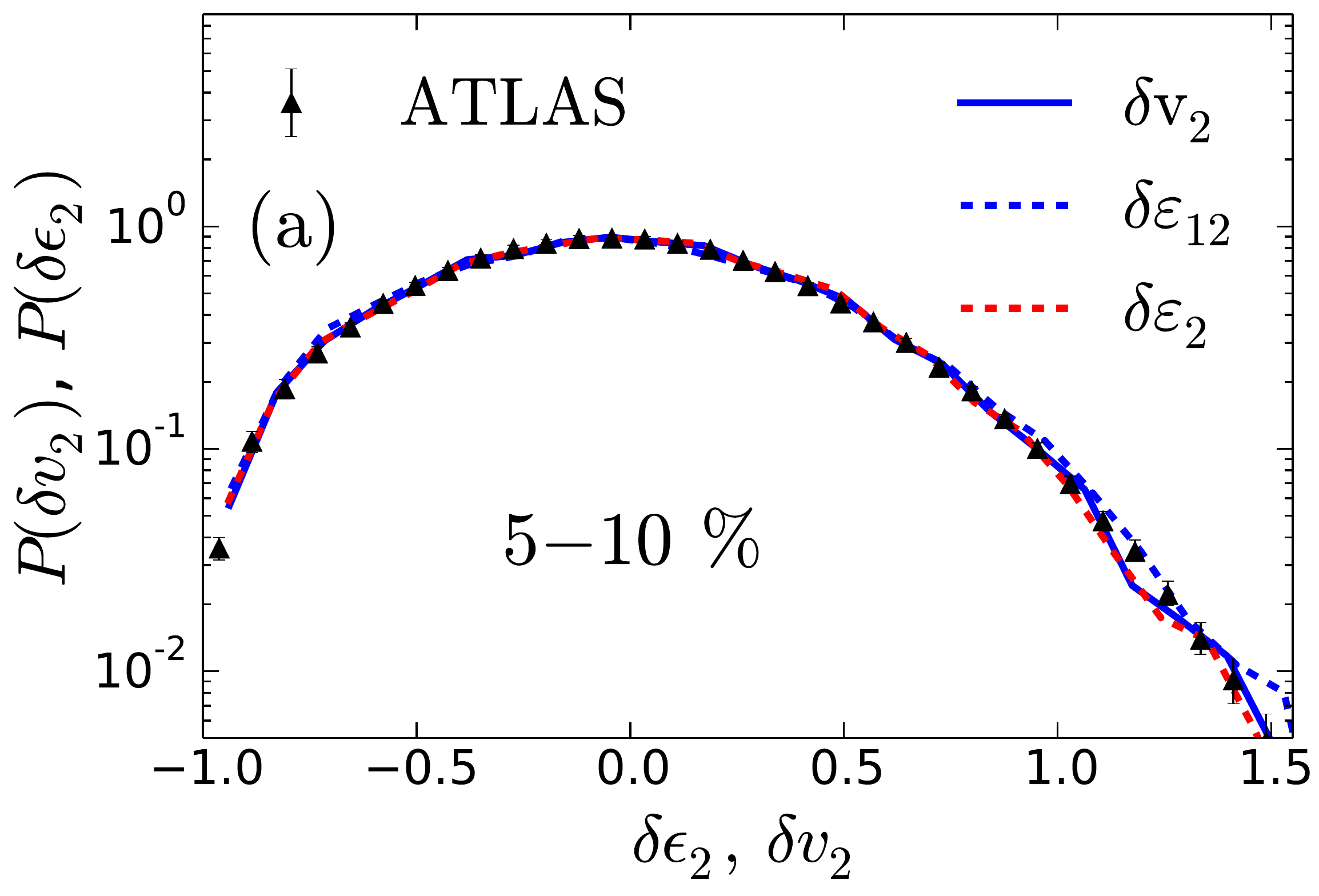}
\includegraphics[width=5.5cm]{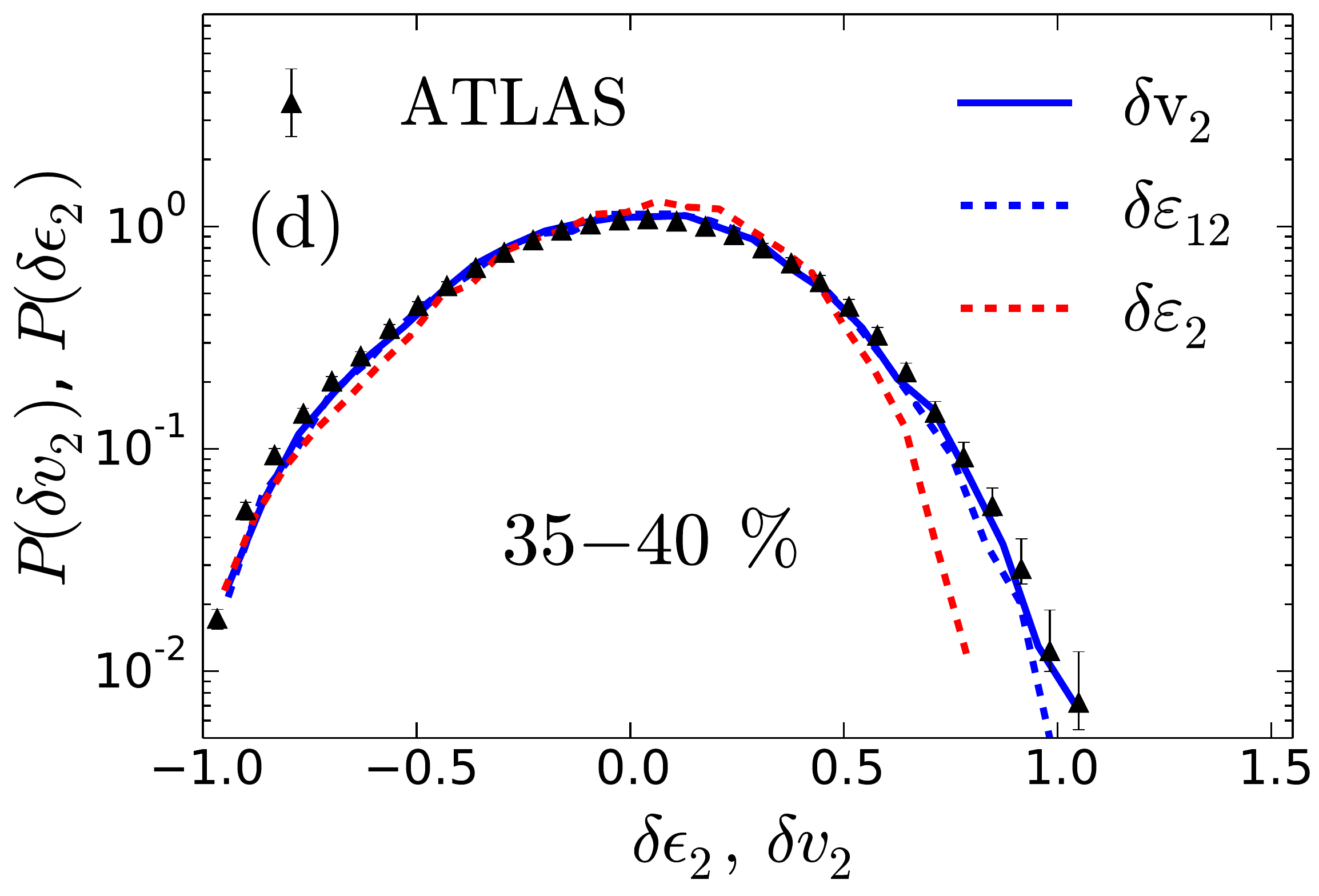}
\includegraphics[width=5.5cm]{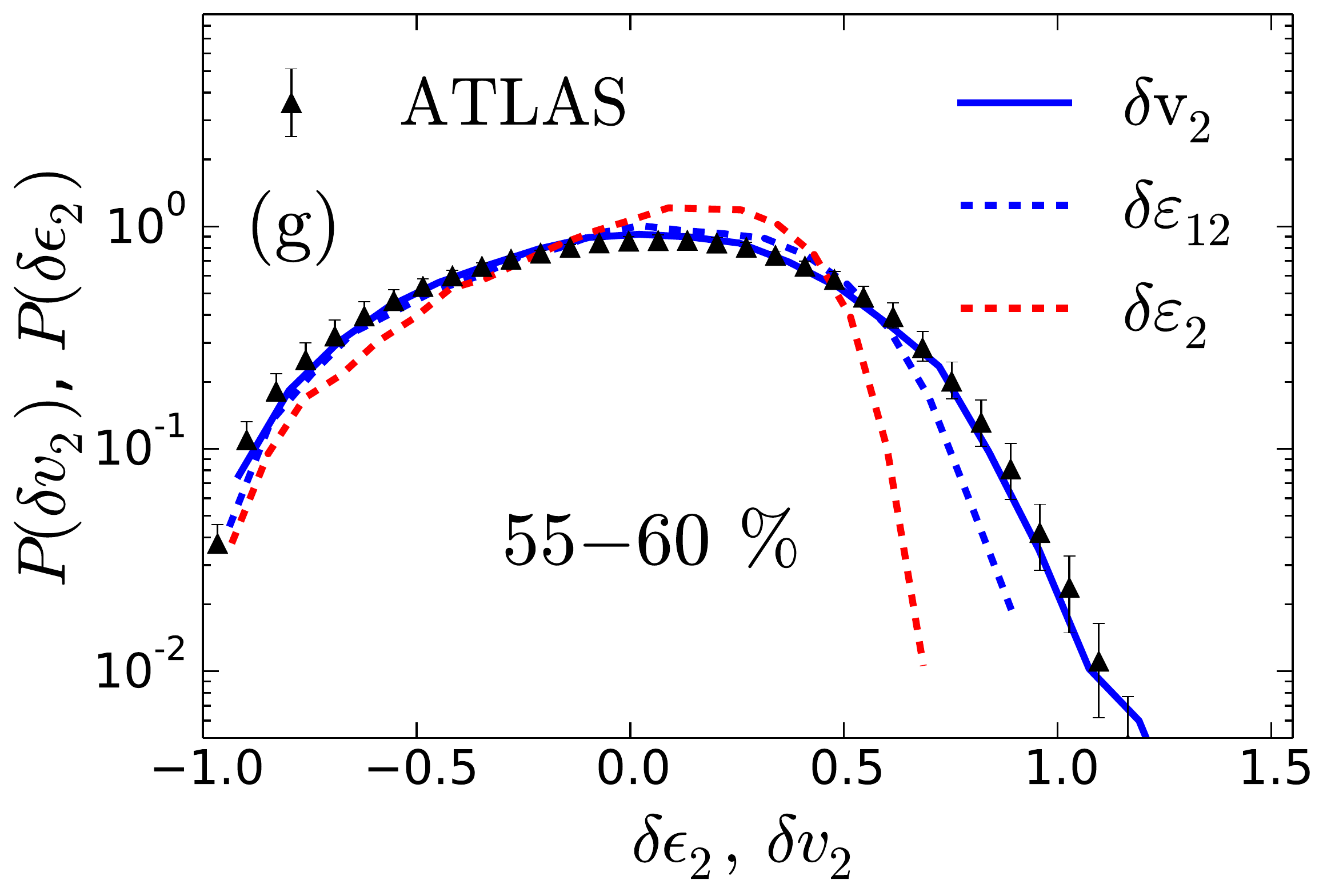}
\vspace{-0.7cm}
\caption{\small Probability distributions of the charged hadron $\delta v_2$, and of the initial $\delta \varepsilon_{2}$ (and $\delta \varepsilon_{1,2}$, see \cite{Niemi:2015qia}) in three different centrality classes in $\sqrt{s_{NN}} = 2.76$  TeV Pb+Pb collisions at the LHC. Figures from \cite{Niemi:2015qia}, and experimental data from ATLAS \cite{Aad:2013xma}.}
\label{fig:deltav2dist}
\vspace{-0.2cm}
\end{figure*}

Figure \ref{fig:deltav2dist} shows the probability distributions of the charged hadron
$\delta v_2 = (v_2 - \langle v_2\rangle_{\rm ev})/\langle v_2\rangle_{\rm ev}$ 
and the initial spatial eccentricities $\delta \epsilon_2 = (\epsilon_2 - \langle \epsilon_2\rangle_{\rm ev})/\langle \epsilon_2\rangle_{\rm ev}$ at the LHC in different centrality classes. Our EbyE framework catches remarkably well the centrality systematics of the measured
$P(\delta v_n)$, and also proves that since $\delta \epsilon_2$ and $\delta v_2$ are \textit{nonlinearly correlated} in non-central collisions, hydrodynamics (collectivity) is indeed required to reproduce the measured $P(\delta v_n)$. Furthermore, the $P(\delta v_n)$ offer direct constraints for the initial state in that they are \textit{insensitive} to $\eta/s$. 

\begin{figure*}
\includegraphics[width=5.2cm]{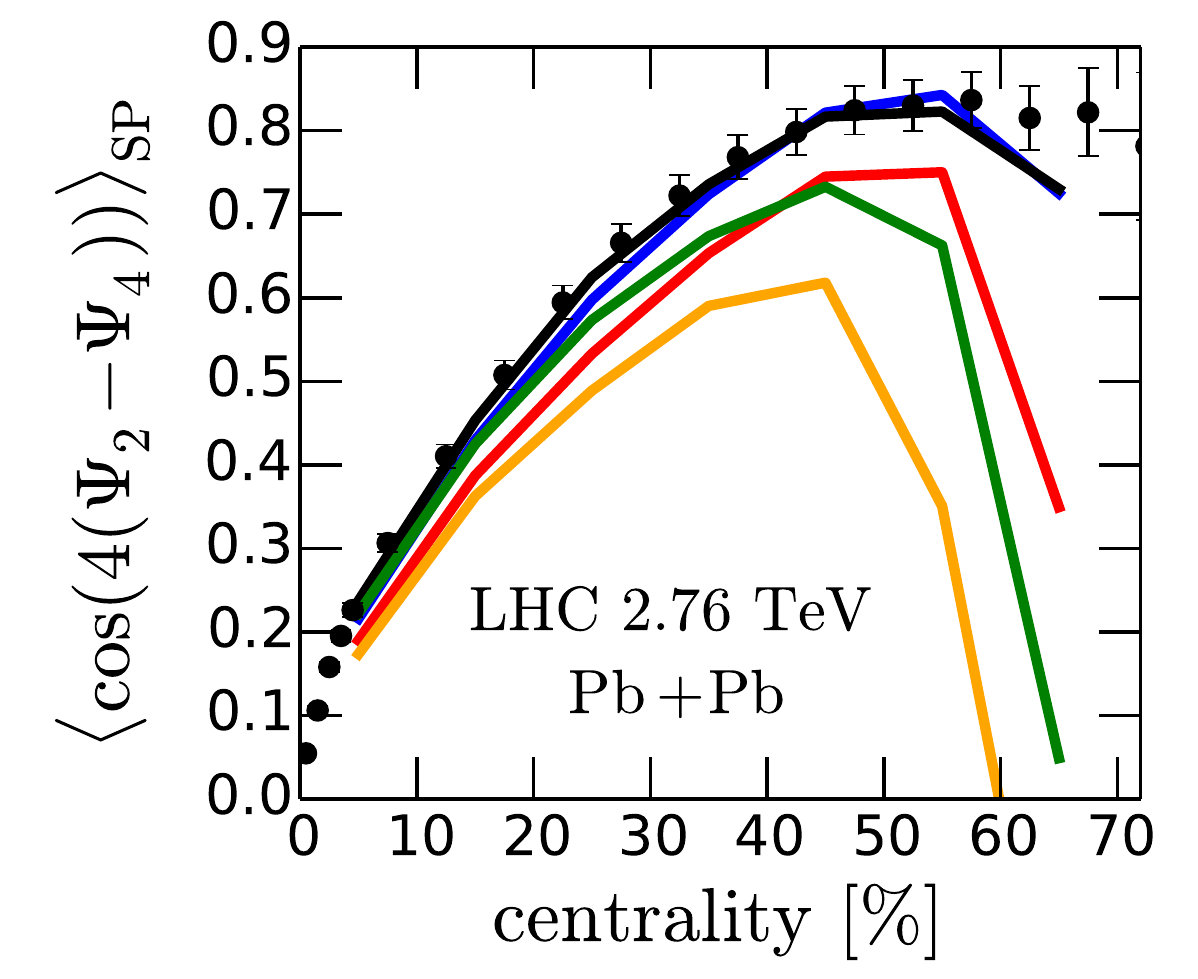} \hspace{-0.3cm}
\includegraphics[width=5.2cm]{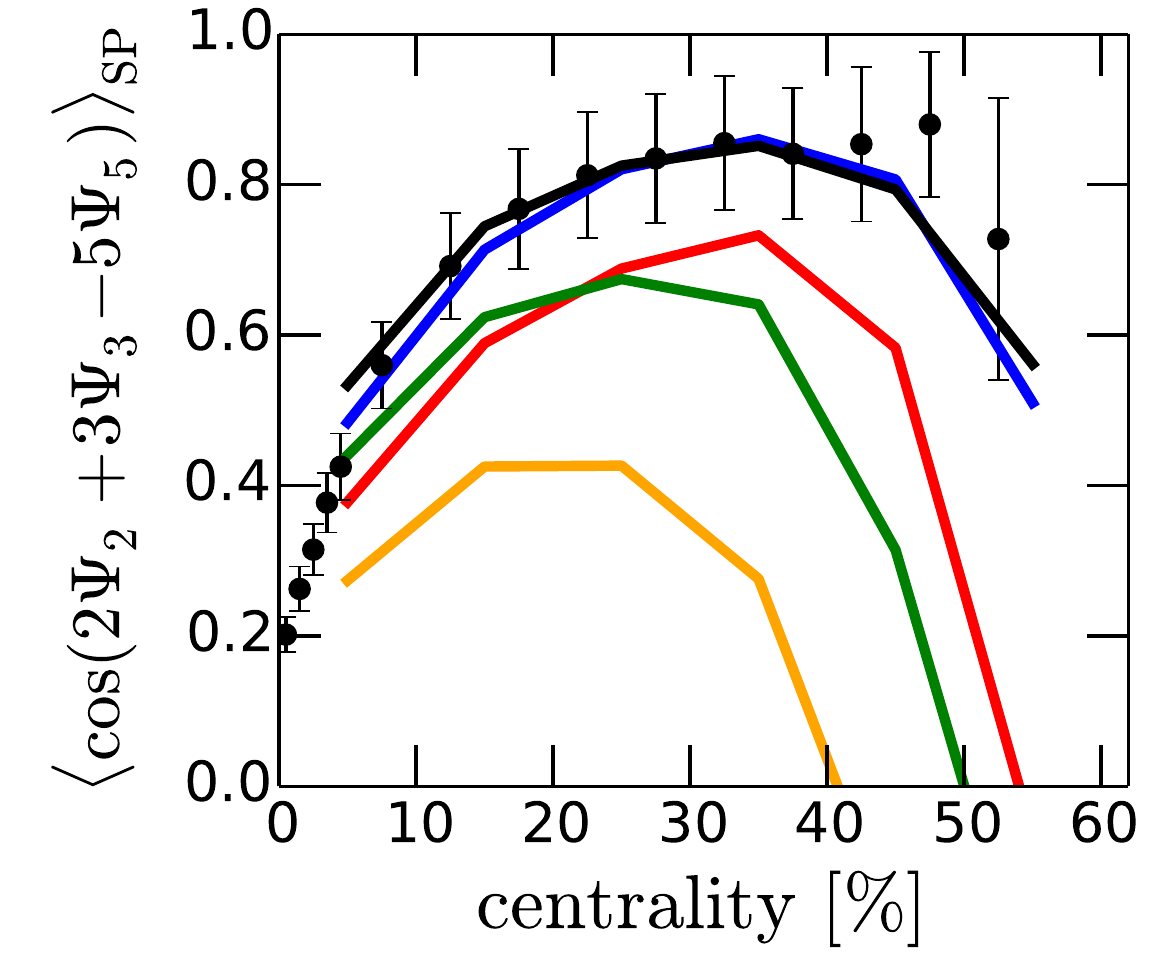}\hspace{0.7cm}
\includegraphics[width=5.2cm]{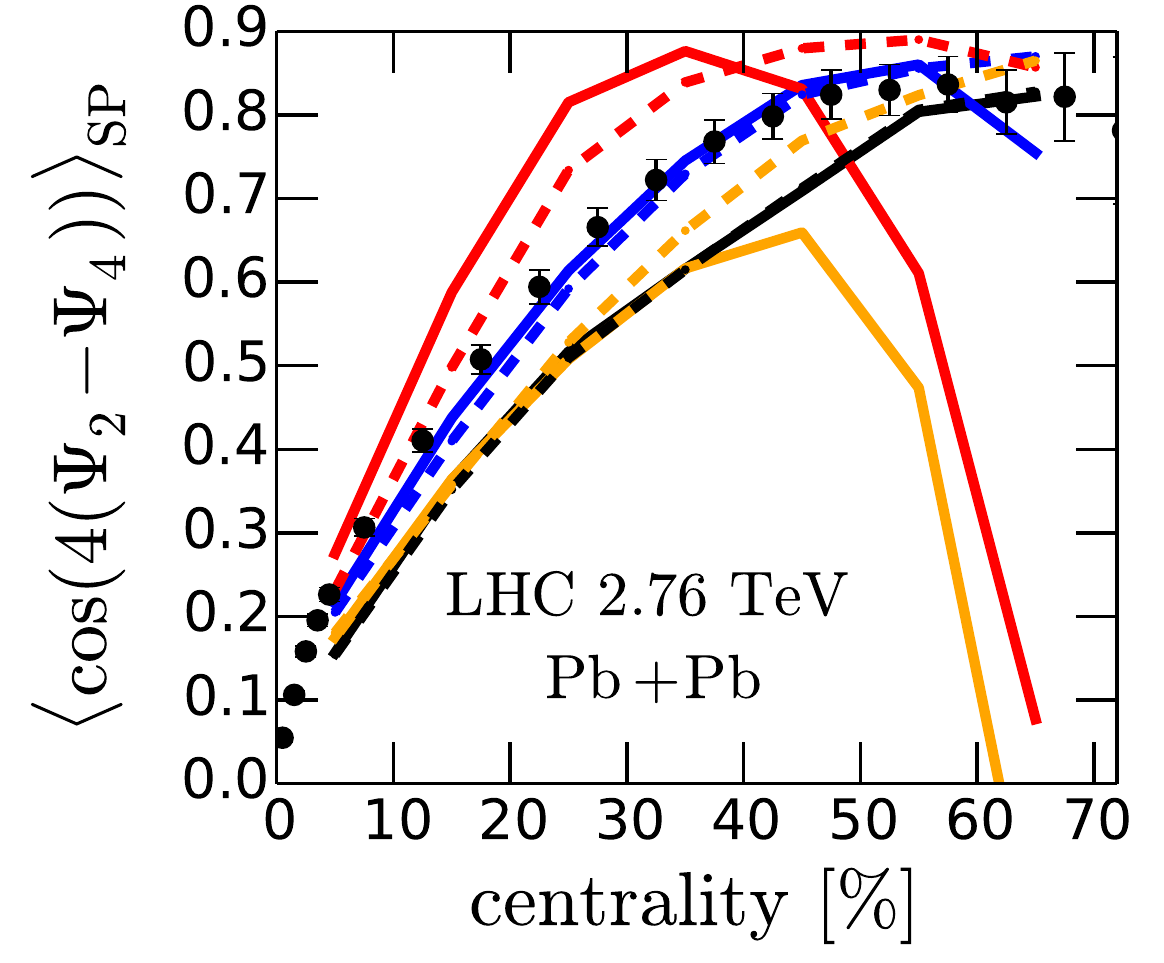}\hspace{-0.3cm}
\vspace{-0.3cm}
\caption{\small Examples of correlations of two (\textit{left}) and three (\textit{middle}) event-plane angles for charged particles at the LHC, compared with the ATLAS data \cite{Aad:2014fla}. Color code as in Fig.~\ref{fig:etapers}.
\textit{Right}: As the left panel but computed for $\eta/s=$ 0.1 (black), 0.2 (blue), 0.3 (red) and $param4$ (orange) with (solid) and without (dashed) $\delta f$ corrections. 
From \cite{Niemi:2015qia}.
}
\label{fig:eventplane_correlation2}
\end{figure*}

Figure \ref{fig:eventplane_correlation2} shows examples of correlations of two and three event-plane angles, $\langle \cos(k_1\Psi_1 + \cdots + n k_n\Psi_n)\rangle_{{\rm SP}}$ at the LHC.  Especially since the $P(\delta v_n)$ now constrain our initial state calculation independently of $\eta/s$, these correlations give vital further constraints for the viscosity and for the validity of the EbyE viscous framework. Remarkably, 
the same two $\eta/s(T)$ parametrizations that explain the RHIC $v_n$’s (black and blue in Fig.~\ref{fig:etapers}) work best also at the LHC! Importantly, as the rightmost panel shows, for these cases also the $\delta f$ corrections remain conveniently small from central to semicentral collisions, suggesting that the obtained hydrodynamic results appear trustworthy up to 40-50\% centrality classes. For more data and discussion, consult again Ref.~\cite{Niemi:2015qia}.


\section{Conclusions and outlook}
\label{conclusions}
Performing a simultaneous LHC and RHIC multi-observable analysis, we have shown that the new NLO-improved EbyE EKRT framework explains remarkably consistently the LHC and RHIC bulk observables in URHIC. The framework has clear predictive power in cms-energy and centrality, and it is a promising tool for getting a controlled estimate of the $T$-dependence of the QCD matter shear viscosity, now that enough orthogonal data constraints start to be available.  
Our results favor a QCD-matter $\eta/s(T)$ which is between a constant value 0.2 and a modestly decreasing(increasing) in the hadron gas(QGP) phase. Similar magnitudes have been obtained also for constant $\eta/s$ by other groups \cite{Song:2010mg,Gale:2012rq}. Next, we need a genuine global analysis to extract true statistical error limits to $\eta/s(T)$. Also the effects of bulk viscosity \cite{Ryu:2015vwa}
 should be studied. Work for the EKRT EbyE framework predictions for the forthcoming 5 TeV Pb+Pb run at the LHC is in progress.   







\end{document}